\documentclass[aps,prb,10pt,superscriptaddress,amsmath,amssymb,twocolumn,amsfonts]{revtex4-2}
\usepackage{amsmath,amssymb,graphicx,bm,comment}
\usepackage[colorlinks=true, linkcolor=blue, citecolor=blue, urlcolor=blue]{hyperref}
\usepackage[utf8]{inputenc}
\usepackage[T1]{fontenc}
\usepackage{tikz}
\usetikzlibrary{positioning}

\usepackage{tikz}
\usetikzlibrary{calc}


\begin{document}

\title{Quantum Phases in a Two-Dimensional Generalized interacting SSH Model}

\author{Rahul Ghosh}
\email{rahul.ghosh@niser.ac.in}
\affiliation{School of Physical Sciences, National Institute of Science Education and Research, Jatni 752050, India}
\affiliation{Homi Bhabha National Institute, Training School Complex, Anushaktinagar, Mumbai 400094, India}


\begin{abstract}
We study interaction-driven quantum phases in a two-dimensional {\it generalized} Su--Schrieffer-Heeger (SSH) defined on a square lattice with inequivalent nearest-neighbor hopping, next-nearest-neighbor hopping and a staggered on-site potential. In the non-interacting limit, the model hosts either a quadratic band touching (QBT) at the Brillouin-zone center or symmetry-protected Dirac nodes, depending on microscopic parameters. In the parameter regime with a QBT, our self-consistent Hartree–Fock analysis shows that weak to intermediate interactions can spontaneously break time-reversal symmetry and stabilize a quantum anomalous Hall (QAH) insulating phase with a finite Chern number. Interestingly, this QAH phase is found to weakly break lattice-symmetries, leading to a small but finite nematic bond order. This is in contrast to the standard QAH phase in checkerboard lattice, which preserves all lattice symmetries. Additionally, we find an enhanced bond-nematic Dirac semimetallic (BNDS) phase due to asymmetric hopping, which is thought to be absent in the Hartree–Fock approach. In the parameter regimes where QBT splits into two Dirac nodes, the QAH phase survives up to a finite staggered on-site potential. However, as the staggered potential increases, the QAH phase is suppressed while the BNDS phase grows stronger.      
\end{abstract}

\maketitle

\section{Introduction}
The study of interaction-driven quantum phases are one of the central themes in the field of topological band theory for more than a decades \cite{PhysRevLett.49.405,PhysRevLett.61.2015,PhysRevLett.95.146802,PhysRevLett.95.226801,PhysRevLett.98.106803,RevModPhys.88.021004}. While non-interacting topological band insulators arise from single-particle physics protected by crystalline or time-reversal symmetries \cite{PhysRevLett.95.146802,PhysRevLett.95.226801,PhysRevLett.98.106803}, strong electron–electron interactions can generate topological states in systems that are otherwise trivial. A striking example is the QAH phase, where a quantized Hall response appears without external magnetic fields, but rather from spontaneous breaking of time-reversal symmetry (TRS) \cite{PhysRevLett.100.156401,PhysRevLett.103.046811,PhysRevB.81.085105,PhysRevB.87.085136,PhysRevB.89.165123,PhysRevB.82.115124,PhysRevLett.49.405,vonKlitzing2020quantumHall,PhysRevLett.61.2015,Chang2013quantumAnomalousHall,Deng2020_QAH_MnBi2Te4,PhysRevLett.127.197701,PhysRevLett.93.206602,RevModPhys.82.1539,RevModPhys.82.1959,Chen2021_TopologicalMottInsulator,PhysRevB.103.035427,PhysRevLett.128.157201,PhysRevB.107.L241105,PhysRevA.93.043611,PhysRevA.86.053618}.

An ideal platform to realise such an interaction-induced topological phase is semimetallic band structures exhibiting either Dirac or QBT points. For Dirac semimetals, mean-field analyses~\cite{PhysRevLett.100.156401,PhysRevB.81.085105,PhysRevB.89.165123} suggest that finite interactions can stabilize a QAH state. However, a few counter analytical and numerical studies~\cite{PhysRevB.88.245123,PhysRevB.88.075101,PhysRevB.89.035103,Guo2014,PhysRevB.92.085147,PhysRevB.92.085146,PhysRevB.92.155137} indicate that charge-ordered phases tend to dominate. In contrast, QBTs are shown to be far more susceptible to interaction-driven instabilities even at weak coupling~\cite{PhysRevLett.103.046811,PhysRevLett.113.106401,PhysRevB.82.115124,PhysRevB.77.235125,PhysRevB.78.245122,Tsai_2015}  due to finite density of states at the Fermi level. These instabilities in turn can give rise to a wide range of ordered phases such as QAH, quantum spin Hall (QSH), nematic, and charge-density-wave (CDW) states. The work by Sun \textit{et al.} \cite{PhysRevLett.103.046811} 
and Uebelacker \textit{et al.}\cite{PhysRevB.84.205122} however raises concern on the stability of QBT point of the non-interacting bands once interaction is introduced. Despite the debate on the stability of QBTs, a great volume of work supports the stable quantum phases including QAH in several lattice models such as checkerboard lattice (\textit{et al.}~\cite{PhysRevLett.117.066403}), the 
Kagome lattice~\cite{PhysRevLett.104.196401,PhysRevB.82.075125,PhysRevLett.117.096402,PhysRevA.98.023609,PhysRevB.98.205146,PhysRevX.14.021046} and bilayer graphene~\cite{Yan2011BilayerGraphene,PhysRevLett.109.126402,PhysRevLett.117.086404,PhysRevB.98.245128,PhysRevB.104.045101}. Thus QBTs are known to provide an ideal setting for probing interaction effects.



While the existence of a QAH phase in systems with QBTs or Dirac nodes remains debated, it is clear that additional competing orders can emerge, depending strongly on the numerical methods employed. For instance, bond-nematic phases do not appear in mean-field Hartree-Fock calculations but can be captured using quantum Monte Carlo \cite{liu2025interaction} simulations and density matrix renormalization group (DMRG)~\cite{Zeng2018_tuning,PhysRevB.109.L081106}. Even when more advanced numerical methods detect such an order, the region of this phase remains very narrow in the parameter space. In view of these and the limitations of lattice models in hosting QAH and bond-nematic orders, we aim to address some fundamental questions: Can Hartree-Fock detect bond-nematic order in a 2D lattice when lattice symmetry is tuned? Can a single lattice model host both QBTs and Dirac nodes, allowing us to study the stability of the QAH phase as the system evolves from QBT to Dirac nodes? can bond-nematic order itself be strengthened by varying lattice parameters?

 In this work, we show that the two-dimensional (2D) generalized SSH model \cite{PhysRevLett.118.076803} with staggered potential provides a minimal platform that can answer to all the above questions in the presence of density-density interaction. In the non-interacting limit, the spectrum hosts either Dirac points or a QBT at the Brillouin-zone center, depending on the choice of parameters. While it shares square symmetry and QBT features with the checkerboard lattice, it differs in its unit-cell connectivity. This allows for distinct symmetry-permitted interaction-driven phases. Using unrestricted Hartree-Fock approach, we chart out the ground state phase diagram at half-filling in the weak-to-intermediate-coupling regime. Our analysis reveals a rich variety of interaction-driven ordered phases, including a spontaneous QAH phase characterized by topological invariant and charge-ordered states that break lattice translational symmetry. We also identify a BNDS, absent in earlier Hartree-Fock studies ~\cite{PhysRevLett.125.240601,PhysRevB.106.235103}, which appears between the QAH and site-nematic insulating (SNI) phases. The stability region of this phase expands significantly in the presence of asymmetric hopping. We further investigate the stability of the QAH phase under a staggered on-site potential, which drives the noninteracting system into a Dirac semimetallic phase. Remarkably, we find that interactions can also induce a QAH state in this regime, adiabatically connected to its QBT-derived counterpart. 
 



\begin{figure}[ht]
	\centering
	\includegraphics[width=1.0\linewidth]{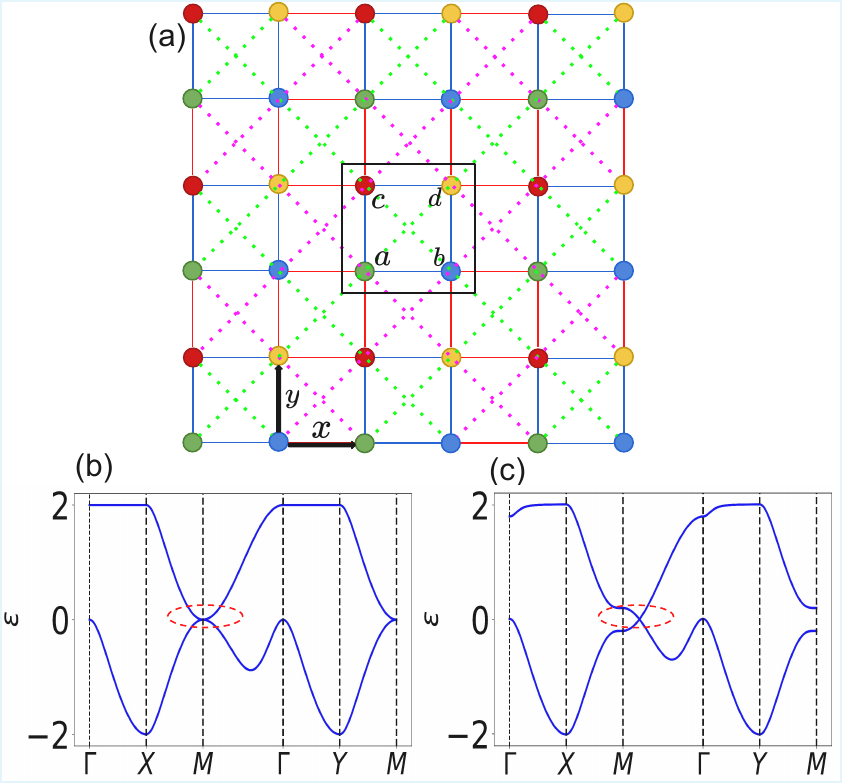}  
	\caption{Lattice geometry and noninteracting band structure of the generalized SSH model.  
    \textbf{(a)} Schematic of the two-dimensional Su–Schrieffer–Heeger (SSH) model on a square lattice with four sites per unit cell (plaquette). Nearest-neighbor (NN) hoppings include intra-cell bonds (blue, amplitude \(t_1\)) and inter-cell bonds (red, amplitude \(t_2\)). Next-nearest-neighbor (NNN) hoppings introduce diagonal couplings: green and purple bonds denote amplitudes \(t_1'\) and \(t_2'\), respectively.  
    \textbf{(b)} Noninteracting band structure at \(\delta=0\), showing a quadratic band touching (QBT) at the Brillouin-zone M point \((\pi,\pi)\).  
    \textbf{(c)} Band structure at \(\delta=-0.2t_1\) with \(t_1=1.0\) and \(t_2=2.0\), where the QBT splits into two Dirac cones. The high-symmetry points are \(\Gamma=(\tfrac{\pi}{2},\tfrac{\pi}{2})\), \(X=(\pi,\tfrac{\pi}{2})\), \(Y=(\tfrac{\pi}{2},\pi)\), and \(M=(\pi,\pi)\).}
    \label{fig:SSH}
\end{figure}


\section{Model and method}\label{sec1}
We consider a two-dimensional (2D) interacting spinless fermions model Hamiltonian defined on a generalized SSH lattice embedded in a square geometry as shown in Fig.~\ref{fig:SSH}(a). The model features a four-site unit cell with inequivalent intra-cell and inter-cell hopping amplitudes, extended density-density interactions, and a staggered on-site potential. With this, the Hamiltonian reads off
\begin{equation}
\begin{aligned}
\hat{\mathcal{H}} = &- \sum_{i,j} \left( t_{ij} c_i^\dagger c_j + \text{h.c.} \right)
+ \delta \sum_{i \in \{a,d\}} c_i^\dagger c_i 
- \delta \sum_{i \in \{b,c\}} c_i^\dagger c_i \\
&+ U \sum_{\langle i,j \rangle} n_i n_j 
+ V \sum_{\langle\langle i,j \rangle\rangle} n_i n_j ,
\end{aligned}
\label{eq:Hamiltonian}
\end{equation}
where $c_i^\dagger$ ($c_i$) creates (annihilates) a fermion at site $i$, and $n_i = c_i^\dagger c_i$ is the number operator. 
Nearest-neighbor (NN) hoppings [Fig.~\ref{fig:SSH}(a)] are characterized by two amplitudes: $t_{ij}=t_1$, corresponding to intra-cell hopping (blue bonds), and $t_{ij}=t_2$, corresponding to inter-cell hopping (red bonds). In addition, the model includes next-nearest-neighbor (NNN) hoppings, with amplitudes $t_{ij}=t_1'$ (green bonds) and $t_{ij}=t_2'$ (purple bonds).
A staggered potential $\delta$ assigns on-site energies $+\delta$ to sites $a$ and $d$, and $-\delta$ to sites $b$ and $c$.  
Finally, the last two terms account for repulsive density-density interactions: $U$ between NN pairs $\langle i,j\rangle$, and $V$ between NNN pairs $\langle\langle i,j\rangle\rangle$.

Depending on the hopping parameters, the model interpolates between different lattice geometries. For $t_1 = t_2$, the system reduces to a checkerboard lattice with uniform NN hopping. When the NNN hoppings satisfy $t_1' = -t_2' = 0.5$, the model preserves particle-hole symmetry, and the non-interacting band structure exhibits a QBT at the high symmetry $M (\pi,\pi)$ point. A finite staggered potential $\delta$ breaks the fourfold ($C_4$) rotational symmetry and splits the QBT into two Dirac cones~\cite{PhysRevB.106.205105}.  Depending on the sign of $\delta$, the cones appear along one of two orthogonal directions in the Brillouin zone: either the $k_y-k_x=0$ line or the $k_y+k_x=\tfrac{3\pi}{2}$ line. In contrast, setting $t_1 \neq t_2$ breaks particle–hole symmetry but preserves the QBT at $M$ when $\delta=0$ [Fig.~\ref{fig:SSH}(b)]. For finite $\delta$, this QBT also splits into a pair of Dirac cones [Fig.~\ref{fig:SSH}(c)], similar the checkerboard case~\cite{PhysRevB.106.205105}.
The QBT is protected by the $C_4$ rotational symmetry of the lattice, which forbids linear Dirac terms in the effective low-energy Hamiltonian~\cite{PhysRevLett.103.046811, PhysRevB.84.205122}. Breaking $C_4$ symmetry lifts this protection, splitting the QBT into Dirac cones and strongly modifying the susceptibility to interaction-driven instabilities~\cite{PhysRevB.98.125144, PhysRevB.106.205105}.

To study the role of electron-electron interactions, we employ an unrestricted Hartree-Fock (HF) mean-field approximation with a four-site unit-cell ansatz [Fig.~\ref{fig:SSH}(a)]. The density-density interaction in Eq.~(\ref{eq:Hamiltonian}) are decoupled by retaining both Hartree (density) and Fock (exchange) channels, leading to the decomposition  
\begin{equation}
\hat{n}_i \hat{n}_j \simeq 
 \langle \hat{n}_i \rangle \hat{n}_j 
+ \langle \hat{n}_j \rangle \hat{n}_i 
- \langle \hat{n}_i \rangle \langle \hat{n}_j \rangle - \epsilon_{ji} \hat{c}^\dagger_i \hat{c}_j 
- \epsilon_{ij} \hat{c}^\dagger_j \hat{c}_i 
+ |\epsilon_{ij}|^2 
.
\label{HFdecoupling}
\end{equation}
Here, $\langle n_i \rangle$ describes local density modulations, while $\epsilon_{ij}=\langle c_i^\dagger c_j \rangle$ accounts for nonlocal bond-order correlations. We solve the mean-field Hamiltonian on finite lattices with periodic boundary conditions in momentum space by numerical diagonalization. The self-consistant equations of the HF are iteratively updated for the mean field parameters $\langle \hat{n}_{i} \rangle$, until the total energy converges within a tolerance of $10^{-10}$.  All calculations are carried out at a fixed temperature $T=10^{-5}$ (in units of the hopping amplitude), which effectively corresponds to the zero-temperature limit since thermal effects are negligible. The detailed construction of the Hartree-Fock Hamiltonian is presented in Appendix~\ref{A}. From this analysis, we find four distinct competing ordered phases: the site-nematic insulator (SNI), the bond-nematic Dirac semimetal (BNDS), the stripe charge-density-wave (CDW) insulator, and the interaction-driven quantum anomalous Hall (QAH) phase, as will be discussed in the following sections.

\section{Phase diagram in quadratic band touching regime}\label{III}
\begin{figure*}[t]
    \centering
    \includegraphics[width=0.9\textwidth]{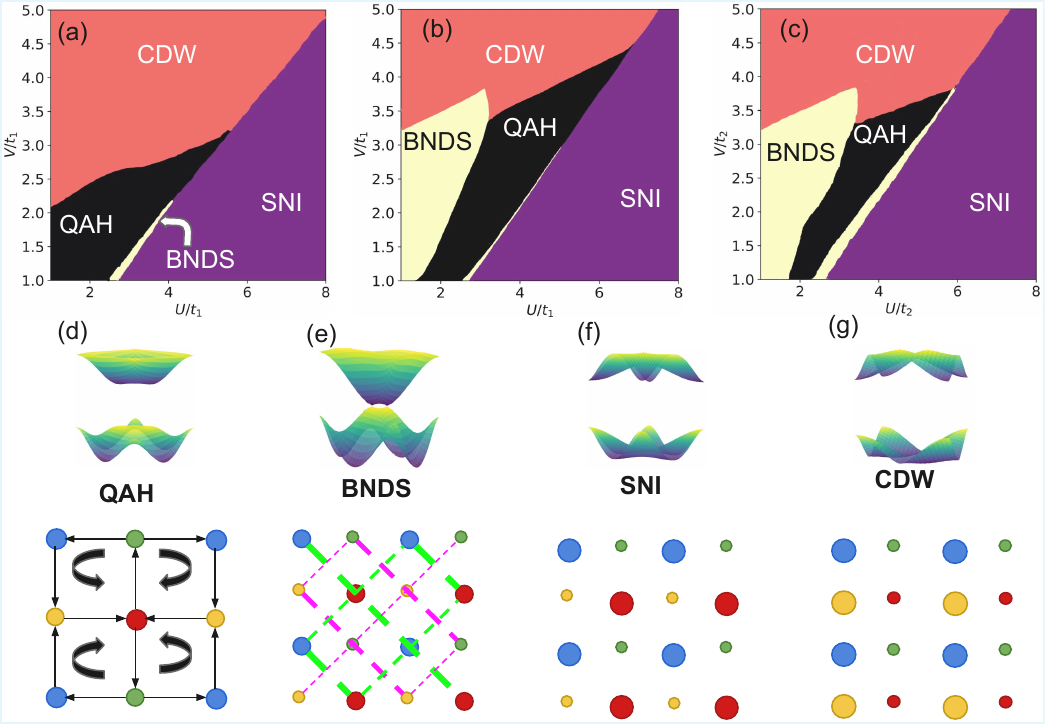} 
    \caption{
    \textbf{Interaction-driven phase diagrams of the generalized 2D SSH model at $\delta = 0$ for different hopping configurations.}
    \textbf{(a)} Phase diagram for the symmetric hopping case $t_1 = t_2 = 1.0$, corresponding to the checkerboard limit. Four distinct phases emerge: a QAH phase with spontaneous time-reversal symmetry breaking (black), a SNI that breaks $C_4$ symmetry (purple), a stripe CDW (red), and a BNDS with gapless anisotropic Dirac nodes (yellow).
    \textbf{(b)} Phase diagram for asymmetric hopping with $t_1 = 1.0$ and $t_2 = 2.0$, representative of the generalized SSH regime. The QAH phase remains robust, while the BNDS region expands significantly due to enhanced bond anisotropy. The site-nematic and stripe-ordered insulating phases persist but shrink in extent. 
    \textbf{(c)} Phase diagram for another asymmetric case, $t_1 = 2.0$ and $t_2 = 1.0$. While the overall structure is similar to (b), the QAH region is reduced, reflecting the sensitivity of topological ordering to the direction of hopping asymmetry. The BNDS and insulating phases continue to dominate, underscoring their robustness. \textbf{(d-g)} Representative band spectra and ordering patterns of spinless fermions at half-filling in the topologically trivial SSH lattice with a QBT point, shown for asymmetric hopping parameters $t_1 = 1.0$ and $t_2 = 2.0$. The QAH phase is distinguished by loop-current order, while the BNDS phase features anisotropic bond ordering. SNI and stripe  CDW phases are also illustrated with their characteristic charge distributions. Circle sizes indicate the local charge density, and bond widths represent the magnitude of the bond-order parameter.
    }
    \label{fig:phase_diagrams}
\end{figure*}

Figure~\ref{fig:phase_diagrams} shows the zero-temperature HF phase diagrams of the generalized 2D SSH model  at half filling with $\delta=0$ for three representative hopping configurations: (a) the symmetric limit $t_1 = t_2 = 1.0$ corresponding to the checkerboard lattice, (b) an asymmetric case with $t_1 = 1.0$, $t_2 = 2.0$, representative of a prototypical 2D SSH regime, and (c) another asymmetric case with $t_1 = 2.0$, $t_2 = 1.0$.


In the symmetric checkerboard limit [Fig.~\ref{fig:phase_diagrams}(a)], our analysis reveals four distinct many-body phases, arising from different mechanisms of symmetry breaking. The QAH insulator (black) breaks time-reversal symmetry via interaction-induced imaginary nearest-neighbor hopping, characterized by the order parameter $\epsilon_{\mathrm{QAH}}$ [Eq.~\eqref{eq:QAH}] and a Chern number $C = \pm 1$~\cite{PhysRevLett.125.240601}.
 In contrast, the SNI (purple) opens a gap while breaking the underlying $C_4$ rotational symmetry of the lattice, a feature captured by the order parameter $\delta_{\mathrm{SNI}}$ [Eq.~\eqref{eq:SNI}]. The stripe CDW  (red) manifests itself as a staggered modulation in the charge density across different lattice sites, which is quantified by $\delta_{\mathrm{CDW}}$ [Eq.~\eqref{eq:CDW}]. Finally, the BNDS (yellow) stands out as a gapless phase that retains Dirac nodes yet breaks $C_4$ symmetry through anisotropic next-nearest-neighbor bond ordering. This anisotropy is measured by differences in bond strength along orthogonal directions: specifically, the imbalance of bonds between $a$ and $d$ sublattices is given by $\Delta_{\mathrm{bond}}^1$ [Eq.~\eqref{eq:bond1}], while the corresponding difference between sublattices $b$ and $c$ is described by $\Delta_{\mathrm{bond}}^2$ [Eq.~\eqref{eq:bond2}], and the overall sublattice-dependent anisotropy is captured by the order parameter $\Delta_{\mathrm{bond}}$ in Eq.~\eqref{eq:bond}.

The order parameters characterizing these phases are given by
\begin{align}
    \epsilon_{\mathrm{QAH}} &= \tfrac{1}{4}\,
        \Big|\, \mathrm{Im}\!\left( \epsilon_{ab} + \epsilon_{bd} + \epsilon_{dc} + \epsilon_{ca} \right) \Big|, 
        \label{eq:QAH} \\[6pt]
    \delta_{\mathrm{SNI}} &= \big| (n_a + n_d) - (n_b + n_c) \big|, 
        \label{eq:SNI} \\[6pt]
    \delta_{\mathrm{CDW}} &= |n_a - n_d| + |n_b - n_c|,
        \label{eq:CDW} \\[6pt]
    \Delta_{\mathrm{bond}}^1 &= 
        \Big| \langle c^{\dagger}_{a+\hat{x}+\hat{y}} c_a \rangle \Big| - 
        \Big| \langle c^{\dagger}_{a+\hat{x}-\hat{y}} c_a \rangle \Big|, 
        \label{eq:bond1} \\[6pt]
    \Delta_{\mathrm{bond}}^2 &= 
        \Big| \langle c^{\dagger}_{b+\hat{x}+\hat{y}} c_b \rangle \Big| - 
        \Big| \langle c^{\dagger}_{b+\hat{x}-\hat{y}} c_b \rangle \Big|, 
        \label{eq:bond2} \\[6pt]
    \Delta_{\mathrm{bond}} &= 
        \Big| \, \big| \Delta_{\mathrm{bond}}^1 \big| - \big| \Delta_{\mathrm{bond}}^2 \big| \, \Big| .
        \label{eq:bond}
\end{align}
 The BNDS phase which is absent in earlier HF studies~\cite{PhysRevLett.125.240601,PhysRevB.106.235103}, emerges in our fully unrestricted Hartree-Fock treatment as a robust intermediate phase between the QAH and SNI states in the checkerboard model. This finding is consistent with recent thermodynamic studies~\cite{PhysRevB.109.L081106,liu2025interaction}, which identified BNDS as a finite-temperature stable phase with characteristic bond-order signatures. Our results therefore demonstrate that BNDS ubiquitously intervenes between SNI and other interaction-driven insulating states in the checkerboard lattice, highlighting the richness of the phase diagram in extended Hubbard models on semimetallic lattices.

Upon introducing the hopping asymmetry, as shown in Fig.~\ref{fig:phase_diagrams}(b) for $t_1 = 1.0$, $t_2 = 2.0$, the phase diagram changes markedly. The BNDS phase expands significantly, occupying much of the weak-to-intermediate-coupling regime ($U/t_1-V/t_1$). This behavior arises from an enhanced bond anisotropy due to the hopping imbalance which stabilizes nematic textures even in the semimetallic regime. The QAH phase (black), characterized by a Chern number $C = \pm 1$ [see Appendix \ref{B}], remains present but occupies a reduced region of the phase diagram,  while the stripe and site-nematic insulating phases persist only in restricted regions of the parameter space. The corresponding band structure and representative bond and charge order patterns are shown in Fig.~\ref{fig:phase_diagrams}(d)-~\ref{fig:phase_diagrams}(g).
In the another asymmetric case [Fig.~\ref{fig:phase_diagrams}(c)] with $t_1 = 2.0$, $t_2 = 1.0$, a qualitatively similar structure is observed, confirming the robustness of the BNDS phase under hopping asymmetry. Interestingly, the QAH region is further suppressed as a result of the inversion of the hopping hierarchy, suggesting a sensitive dependence of the topological gap formation on the dominant kinetic pathways. Overall, the BNDS phase remains prevalent across all three hopping regimes, indicating its generic presence in the phase diagram of generalized SSH models with extended interactions.
Our results not only extend previous Hartree-Fock analyses~\cite{PhysRevB.106.235103} but also establish a clear link between bond anisotropy and the stabilization of gapless nematic phases. The clear identification of the BNDS region in our self-consistent solution highlights the critical role of competing bond-order channels in shaping the phase diagram of semimetallic systems with interaction-driven instabilities.

\begin{figure}
	\centering
	\includegraphics[width=1.0\linewidth]{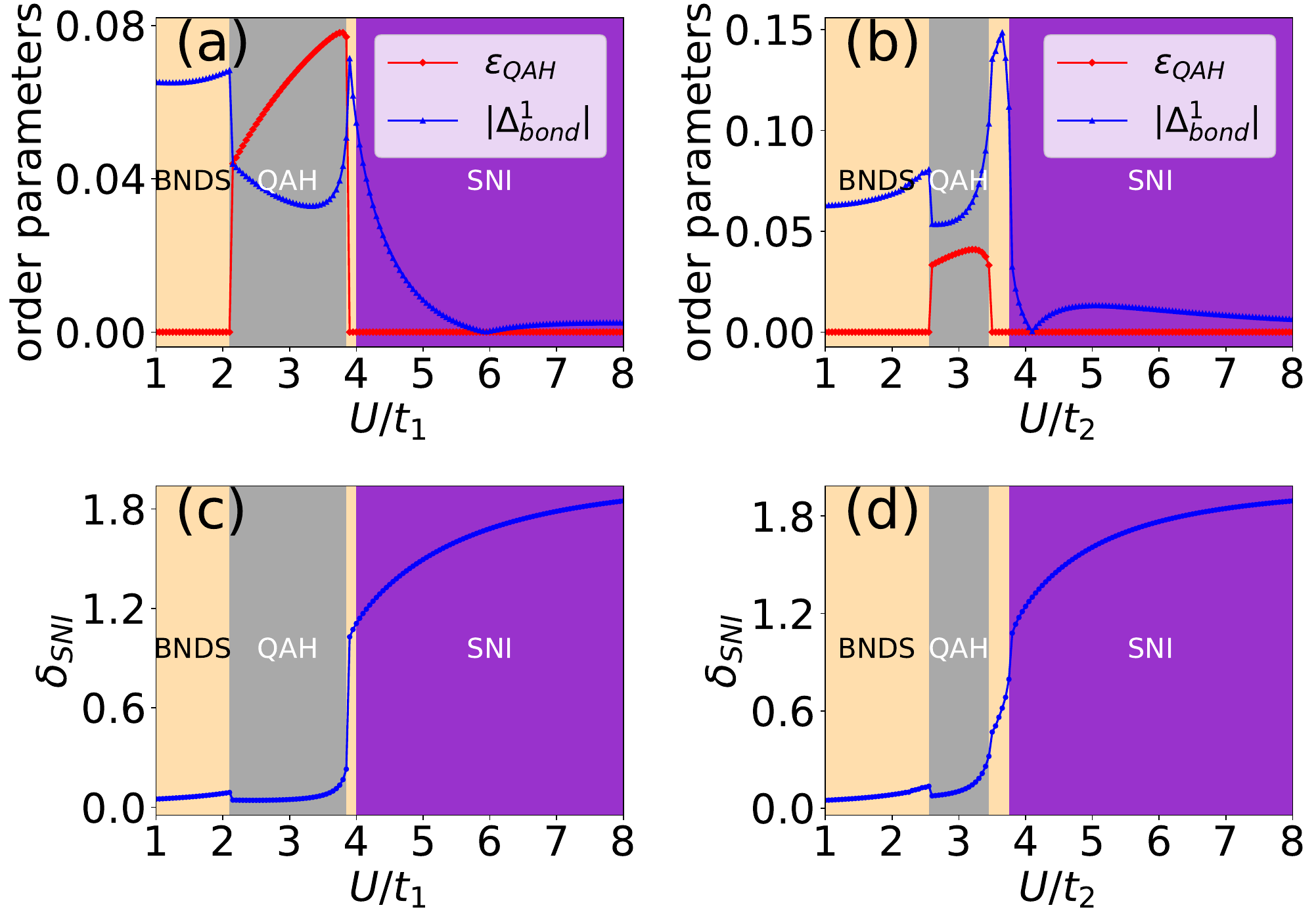}  
	\caption{
    \textbf{(a)} Evolution of the QAH and  BNDS order parameters as functions of the interaction strength \( U \) for fixed hopping amplitudes \( t_1 = 1.0 \), \( t_2 = 2.0 \), and onsite potential \( \delta = 0 \). At weak coupling, the system exhibits a pure BNDS phase characterized by spontaneous bond-nematic order. In the intermediate \( U \) regime, both QAH and BNDS order parameters coexist, indicating an interaction-induced topological-nematic mixed phase. As \( U \) increases further, the QAH order is suppressed, and a narrow BNDS region emerges before the transition to a SNI phase. 
    \textbf{(b)} Corresponding plot for \( t_1 = 2.0 \), \( t_2 = 1.0 \), showing a reduced extent of the QAH phase and a slightly broader BNDS region separating the QAH and SNI phases. All the data in (a) and (b) are in artibitary unit.
    \textbf{(c)} Site-nematic order parameter \( \delta_{\text{site}} \) as a function of \( U \) for \( t_1 = 1.0 \), \( t_2 = 2.0 \). The SNI order parameter remains finite across all phases, including the BNDS and QAH regimes, but becomes strongly enhanced deep in the SNI phase, signaling a crossover from weak to strong site-nematic order. 
    \textbf{(d)} Similar behavior of \( \delta_{\text{site}} \) for \( t_1 = 2.0 \), \( t_2 = 1.0 \), consistent with the trends observed in panel (c), confirming the robustness of site-nematic fluctuations across the phase diagram. All plots are obtained for fixed next nearest-neighbor interaction strength \( V = 2.0t_1 \) in panels (a) and (c), and \( V = 2.0t_2 \) in panels (b) and (d).
    }
    \label{fig:SSH1}
\end{figure}

\section{Order parameters}

Figure~\ref{fig:SSH1} illustrates the evolution of competing and coexisting order parameters as a function of interaction strength \( U \) for two representative sets of hopping asymmetries in the generalized two-dimensional SSH model. We focus on the QAH,  BNDS, and SNI order, each of which breaks distinct symmetries and characterizes different interaction-driven phases.

\subsection{Bond Nematic Dirac Semimetal}
\begin{figure}[ht]
	\centering
	\includegraphics[width=1.0\linewidth]{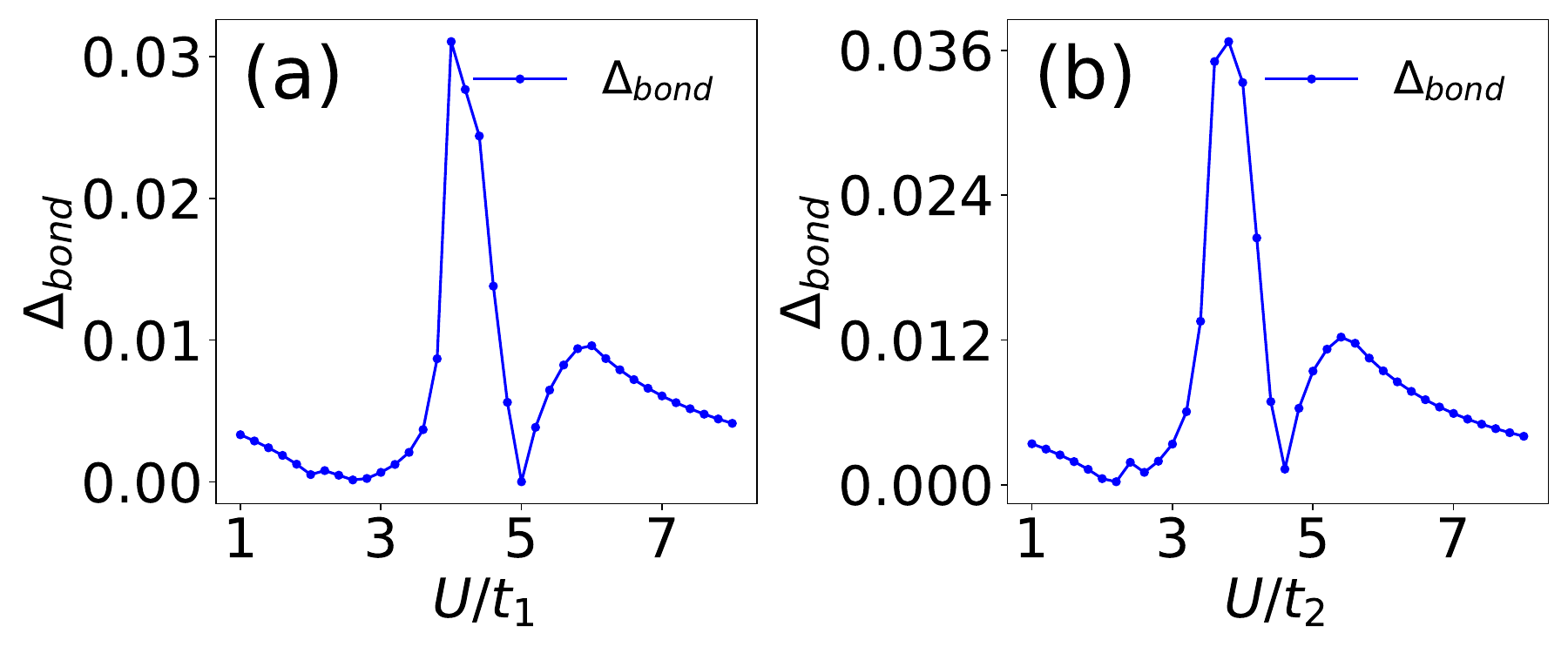}  
	\caption{
	Interaction dependence of the sublattice-dependent bond order.  
	\textbf{(a)} $t_1=1.0,t_2=2.0$ and \textbf{(b)} $t_1=2.0,t_2=1.0$, both exhibiting a finite value that signals sublattice-selective bond order driving anisotropic BNDS formation.  
	For comparison, the symmetric checkerboard limit $(t_1=t_2=1.0)$ yields zero difference, confirming the absence of sublattice dependence.
	}
    \label{fig:anisotropy}
\end{figure}

In Fig.~\ref{fig:SSH1}(a), for the case \( t_1 = 1.0 \), \( t_2 = 2.0 \), and \( \delta = 0 \), we identify three qualitatively distinct regimes as \( U \) increases. At weak coupling, only the bond-nematic order parameter 
\(\Delta_{\mathrm{bond}}^1\) is finite, indicating spontaneous breaking  of \(C_4\) rotational symmetry while time-reversal and inversion symmetries  remain intact, resulting in a vanishing QAH order parameter. This corresponds to a BNDS phase, where the quadratic band-touching point splits into two  Dirac cones as a result of interaction-induced directional modulation of bond amplitudes. Notably, this BNDS phase, emerging from asymmetric hopping in the SSH model, is novel in character: the bond order between the \(a\) and \(d\) sites differs slightly from that between the \(c\) and \(b\) sites, as illustrated in Fig.~\ref{fig:anisotropy}. This anisotropic sublattice-dependent bond order is absent in the checkerboard limit.

\subsection{Coexistence of Bond Nematic Dirac Semimetal and QAH phase}
At intermediate interaction strengths, both  $\epsilon_{QAH}$ and  $\Delta_{\mathrm{bond}}^1$  acquire finite values simultaneously. This co-existence region [Fig.~\ref{fig:SSH1}(a)] corresponds to an \emph{interaction-driven topological-nematic phase}, in which the system simultaneously breaks time-reversal symmetry (via QAH ordering) and rotational symmetry (via bond-nematicity). Interestingly, the emergence of the QAH order tends to suppress the bond-nematic order, indicating a competition between topological and nematic tendencies. While the QAH order opens a topological gap at the QBT, the bond-nematic order preserves the underlying anisotropy in the electronic structure. This regime is particularly noteworthy, as it demonstrates that the QAH phase does not arise from a purely topological instability, but instead emerges together with a preexisting nematic distortion. 


\subsection{Site Nematic Order parameter}
Even more intriguingly, the \emph{site-nematic order parameter} \( \delta_{\text{SNI}} \) as shown in Fig.~\ref{fig:SSH1}(c), remains finite across the entire range of interaction strength \( U \), including within the BNDS and QAH regimes. Although it is subdominant in magnitude in these intermediate phases, \( \delta_{\text{SNI} }\) exhibits a sharp enhancement at large \( U \), signaling a transition into a SNI phase that breaks the \( C_4 \) rotational symmetry and opens a topologically trivial gap. Notably, both the BNDS and SNI phases break the same point-group symmetry of the lattice~\cite{PhysRevB.109.L081106}. 

Fig.~\ref{fig:SSH1}(b) and \ref{fig:SSH1}(d) show the corresponding results for the inverted hopping configuration \( t_1 = 2.0 \), \( t_2 = 1.0 \). Although the qualitative sequence of phases remains similar, there are notable differences: the QAH regime is narrower, while the intermediate BNDS region between the QAH and SNI phases is somewhat broader. This asymmetry reflects the role of hopping anisotropy in stabilizing or suppressing the QAH phase. Specifically, since the QAH state relies on complex loop currents and inter-site coherence, it is more sensitive to bond inhomogeneities introduced by strongly asymmetric hopping. Conversely, nematic phases, which rely on real-space symmetry breaking, are comparatively more robust.

\begin{figure*}
    \centering
    \includegraphics[width=0.9\textwidth]{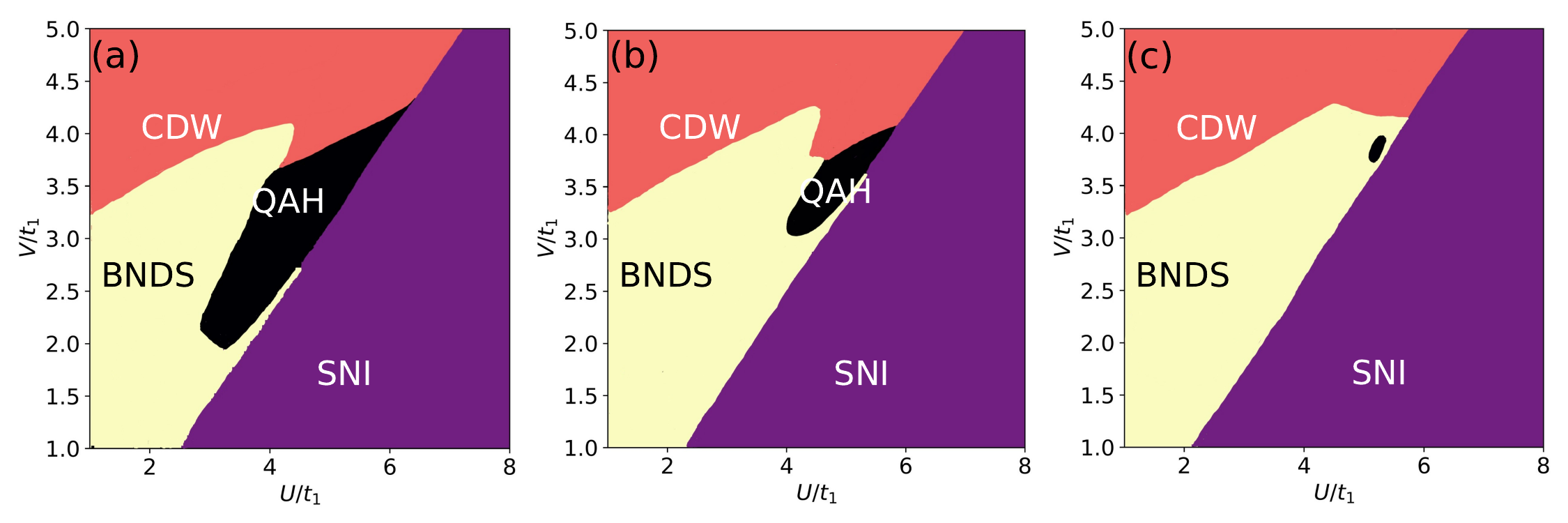} 
    \caption{
    \textbf{Interaction-driven phase diagrams in the Dirac semimetal regime with $t_1 = 1.0$ and $t_2 = 2.0$ for varying staggered potential $\delta$.} 
    \textbf{(a)} For $\delta = 0.2t_1$, a finite staggered potential already reduces the extent of the QAH phase (black) compared to the $\delta=0$ case. 
    \textbf{(b)} At $\delta = 0.4t_1$, the QAH region is further suppressed. 
    \textbf{(c)} For $\delta = 0.6t_1$, the QAH phase is nearly absent, giving way to the BNDS phase. 
    Across all cases, increasing $\delta$ destabilizes the interaction-induced QAH state in favor of BNDS, indicating that the QAH phase remains robust only for weak sublattice staggering.
    }
    \label{fig:phase_diagrams_delta}
\end{figure*}

Another key observation is that the SNI order parameter \( \delta_{\text{SNI}} \) shows similar qualitative behavior in both hopping regimes remaining finite across all \( U \), but sharply increasing in the strong-coupling regime. This robustness of site-nematicity further supports the notion that even the topologically non-trivial phases emerge within a nematically distorted background.  All data presented in Fig.~\ref{fig:SSH1} were computed for fixed  next nearest-neighbor interaction strength \( V = 2.0t_1 \) in panels~\ref{fig:SSH1}(a) and \ref{fig:SSH1}(c), and \( V = 2.0t_2 \) in panels~\ref{fig:SSH1}(b) and \ref{fig:SSH1}(d). The choice ensures a consistent comparison across different hopping asymmetries, highlighting the competition and coexistence among QAH, BNDS, and SNI phases.


\section{Phase diagram in the Dirac semimetal regime}
A finite $\delta$ breaks sublattice symmetry and splits the QBT into two Dirac cones, thereby reducing the density of states at the Fermi level and modifying the balance between competing ordered phases.

Figure~\ref{fig:phase_diagrams_delta} summarizes the evolution of interaction-driven phases with increasing $\delta$ for fixed hopping parameters $t_1 = 1.0$ and $t_2 = 2.0$. At $\delta = 0.2t_1$ [Fig.~\ref{fig:phase_diagrams_delta}(a)], the quantum anomalous Hall (QAH) phase, characterized by a nonzero Chern number $C = \pm 1$, still occupies a sizable region of the interaction parameter space, although its extent is already reduced compared to the case $\delta = 0$ [Fig. \ref{fig:phase_diagrams}(b)]. The explicit breaking of sublattice symmetry and simultaneous splitting of the QBT into Dirac points compete with the interaction-driven spontaneous time-reversal symmetry breaking responsible for the QAH state. Increasing $\delta$ to $0.4t_1$ [Fig.~\ref{fig:phase_diagrams_delta}(b)] further strengthens this competition, significantly suppressing the QAH phase in favor of the BNDS phase.  At $\delta = 0.6t_1$ [Fig.~\ref{fig:phase_diagrams_delta}(c)], the QAH phase is almost eliminated, with the BNDS phase occupying most of the phase diagram. This progression highlights the fragility of the interaction-induced QAH state against explicit lattice symmetry-breaking perturbations: while interactions can overcome weak sublattice staggering to stabilize a topological phase, sufficiently large $\delta$ suppresses the low-energy degeneracies required for QAH order and instead enhances nematic ordering tendencies. These results demonstrate that in the Dirac semimetal regime, the QAH phase remains robust only for small $\delta$, beyond which the symmetry broken bond-ordered phases become energetically favorable.

\section{Summary and Conclusion}\label{sec3}

We have studied the interplay between topology and nematicity in the generalized interacting 2D SSH model, controlled by hopping asymmetry, sublattice potential, and interaction strength. Using an unrestricted HF approach, we identify four competing phases: QAH insulator, SNI, stripe CDW insulator, and BNDS.  In the symmetric checkerboard limit, the BNDS is found to be a robust but narrow intermediate state between QAH and SNI, which was absent in earlier HF studies. Increasing hopping asymmetry enlarges the BNDS region but it reduces the stability of the QAH phase. The order parameters reveal a clear sequence: BNDS dominates at weak coupling; at intermediate coupling it co-exists with the QAH phase; the SNI phase  prevails at strong coupling. Site-nematic order persists across all regimes, indicating that topological phases develop on a nematic background. We further find that, with asymmetric hopping, the BNDS phase also develops a sublattice-dependent bond order, absent in the symmetric checkerboard case. A finite sublattice potential $\delta$ further destabilizes the QAH phase by splitting the quadratic band touching into Dirac cones and lowering the density of states. As a result, the QAH phase remains stable only within a narrow parameter window, while nematic phases, particularly the BNDS, prevail over most of the phase diagram.

These results show that the interaction-driven QAH phase is not a separate instability but part of a broader landscape where nematic and topological orders are closely connected. This is directly relevant for experimental platforms such as cold-atom optical lattices, moiré superlattices, and quantum simulators, where hopping asymmetry and sublattice potentials can be tuned with high precision.Future directions include exploring dynamical control through periodic driving or strain, and extending beyond mean-field theory using approaches such as exact diagonalization, DMRG, or functional renormalization group. These methods can test the robustness of the QAH and BNDS phases and can reveal new types of topological–nematic quantum phases.

\section{Acknowledgments}
We acknowledge financial support from the Department of Atomic Energy (DAE), Government of India, under the project Basic Research in Physical and Multidisciplinary Sciences (RIN4001). RG gratefully acknowledges the use of the Virgo cluster, where most of the numerical calculations were carried out. RG also thanks Kush Saha and Subhajyoti Pal for many insightful discussions, comments and  critical reading of the manuscript.

\appendix
\section{\MakeUppercase{Hartree-Fock mean-field of the interacting Hamiltonian (4-site unit cell ansatz)}}\label{A}

In Sec.~\ref{sec1}, we introduced the HF mean-field method. In this section, we present the explicit HF equations used to solve the interacting Hamiltonian in Eq.~\ref{eq:Hamiltonian} at half filling and zero temperature. The interaction term is decoupled using the Wick decomposition [Eq.~\ref{HFdecoupling}], with $\epsilon_{ij}=\epsilon_{ij}^R+i\epsilon_{ij}^I$. The resulting mean-field Hamiltonian is then expressed in momentum space by adopting a four-site unit-cell ansatz [Fig.~\ref{fig:SSH}(a)].  

To incorporate loop-current  associated with imaginary nearest-neighbor (NN) hoppings, we impose the constraints.
\begin{align*}
\langle \hat{c}^\dagger_{b} \hat{c}_{a} \rangle &= \langle \hat{c}^\dagger_{b} \hat{c}_{a+2\hat{x}} \rangle, &
\langle \hat{c}^\dagger_{c} \hat{c}_{a} \rangle &= \langle \hat{c}^\dagger_{c} \hat{c}_{a+2\hat{y}} \rangle, \\
\langle \hat{c}^\dagger_{d} \hat{c}_{c} \rangle &= \langle \hat{c}^\dagger_{d} \hat{c}_{c+2\hat{x}} \rangle, &
\langle \hat{c}^\dagger_{d} \hat{c}_{b} \rangle &= \langle \hat{c}^\dagger_{d} \hat{c}_{b+2\hat{y}} \rangle,
\end{align*}

Together with their Hermitian conjugates. In addition, the mean-field parameters associated with next-nearest-neighbor (NNN) hoppings are constrained to remain real. The resulting HF Hamiltonian takes the form:

\begin{align}
\hat{H} = E_0 + \sum_{\mathbf{k}} 
\Psi^\dagger_{\mathbf{k}} \,
H^{\mathbf{k}}_{\mathrm{HF}} \,
\Psi_{\mathbf{k}},
\end{align}
with
\[
\Psi_{\mathbf{k}} = 
\begin{pmatrix} 
\hat{c}_{\mathbf{k},a} \\ 
\hat{c}_{\mathbf{k},b} \\ 
\hat{c}_{\mathbf{k},c} \\ 
\hat{c}_{\mathbf{k},d} 
\end{pmatrix}.
\]

\vspace{5pt}
The diagonal elements of $H^{\mathbf{k}}_{\mathrm{HF}}$ read:
\begin{align*}
H^{\mathbf{k}}_{00} &= 2U(\bar{n}_{b} + \bar{n}_{c}) + 4V \bar{n}_{d} + \delta, \\
H^{\mathbf{k}}_{11} &= 2U(\bar{n}_{a} + \bar{n}_{d}) + 4V \bar{n}_{c} - \delta, \\
H^{\mathbf{k}}_{22} &= 2U(\bar{n}_{a} + \bar{n}_{d}) + 4V \bar{n}_{b} - \delta, \\
H^{\mathbf{k}}_{33} &= 2U(\bar{n}_{c} + \bar{n}_{b}) + 4V \bar{n}_{a} + \delta.
\end{align*}

The off-diagonal elements are given by
\begin{align*}
H^{\mathbf{k}}_{01} &= -t_1 e^{-k_x} - t_2 e^{k_x} - 2U \epsilon^R_{ab}\cos k_x - 2iV \epsilon^I_{ab}\cos k_x, \\
H^{\mathbf{k}}_{02} &= -t_1 e^{-k_y} - t_2 e^{k_y} - 2U \epsilon^R_{ac}\cos k_y - 2iV \epsilon^I_{ac}\cos k_y, \\
H^{\mathbf{k}}_{13} &= -t_1 e^{-k_y} - t_2 e^{k_y} - 2U \epsilon^R_{bd}\cos k_y - 2iV \epsilon^I_{bd}\cos k_y, \\
H^{\mathbf{k}}_{23} &= -t_1 e^{-k_x} - t_2 e^{k_x} - 2U \epsilon^R_{cd}\cos k_x - 2iV \epsilon^I_{cd}\cos k_x, \\
H^{\mathbf{k}}_{03} &= 2\big(t_1' - V\epsilon^{1}_{ad}\big)\cos(k_x + k_y) + 2\big(t_2' - V\epsilon^{2}_{ad}\big)\cos(k_x - k_y) ,\\
H^{\mathbf{k}}_{12} &= 2\big(t_2' - V\epsilon^{1}_{bc}\big)\cos(k_x + k_y) + 2\big(t_1' - V\epsilon^{2}_{bc}\big)\cos(k_x - k_y).
\end{align*}

Finally, the constant energy shift is:
\vspace{10pt}

$E_{0} = 2U \Big[ 
|\epsilon_{ab}|^{2} + |\epsilon_{cd}|^{2} + |\epsilon_{bd}|^{2} + |\epsilon_{ac}|^{2} 
- (\bar{n}_{a} + \bar{n}_{d})(\bar{n}_{b} + \bar{n}_{c})\Big] \nonumber + 2V \Big[ 
|\epsilon^{1}_{ad}|^{2} + |\epsilon^{2}_{ad}|^{2} + |\epsilon^{1}_{bc}|^{2} + |\epsilon^{2}_{bc}|^{2} 
- 2(\bar{n}_{a}\bar{n}_{d} + \bar{n}_{b}\bar{n}_{c})
\Big].$\\

\section{\MakeUppercase{Chern number and QAH order parameter}}\label{B}

In Sec.~\ref{III}, we have reported that the QAH phase is characterized by a quantized Chern number. Here, we provide an explicit numerical evidence.  

\begin{figure}[ht]
    \centering
    \vspace{10pt}
    \includegraphics[width=0.48\textwidth]{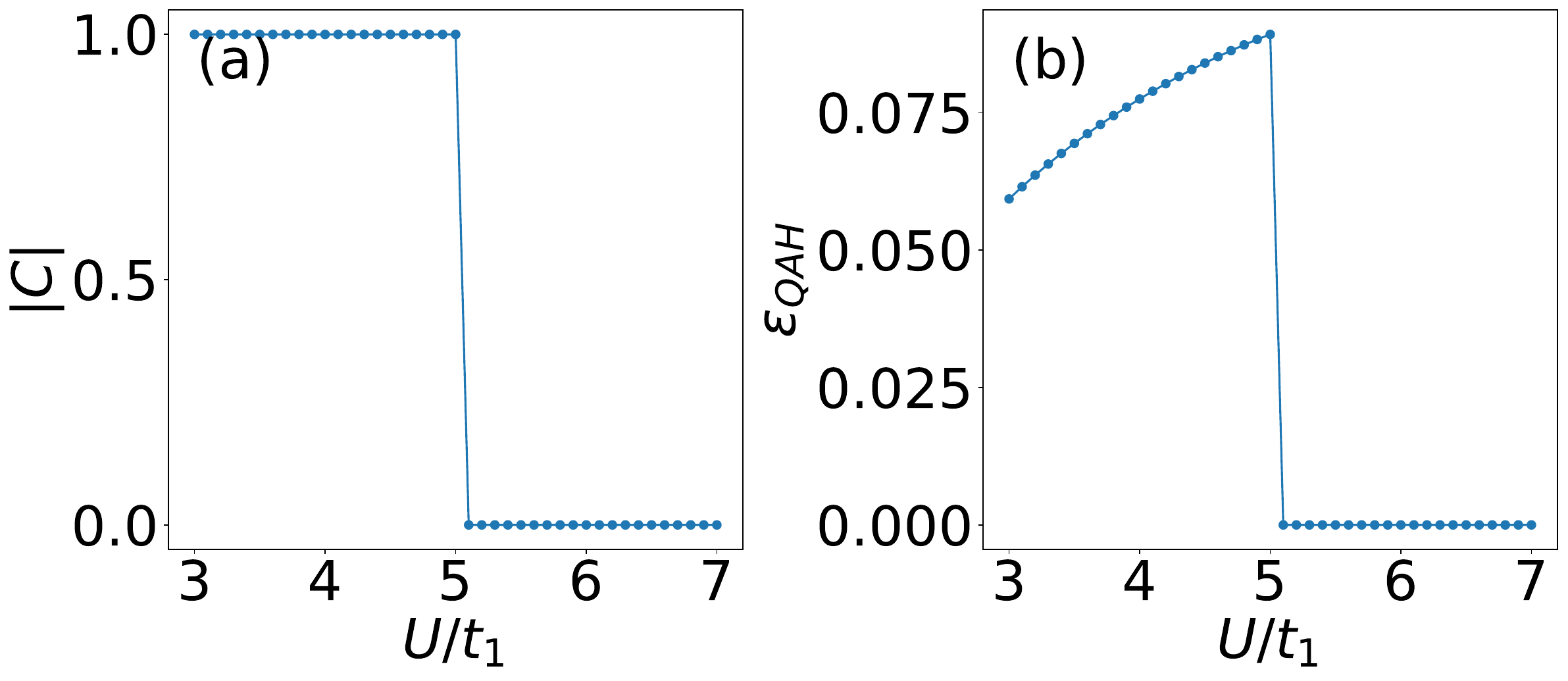} 
    \caption{
    (a) Chern number as a function of the nearest-neighbor interaction strength $U$. 
    The system exhibits a quantized value $|C| = 1$ within the QAH phase. 
    (b) Quantum anomalous Hall (QAH) order parameter as a function of $U$, which is finite in the QAH phase and vanishes upon entering the site-nematic phase. 
    Results are shown for $V = 3.0t_{1}$, $t_{1} = 1.0$, and $t_{2} = 2.0$.
    }
    \label{Chern_no}
\end{figure}

Figure~\ref{Chern_no}(a) shows the dependence of the Chern number on the nearest-neighbor interaction $U$. 
The Chern number of the occupied bands is evaluated using the Fukui--Hatsugai--Suzuki method~\cite{Fukui2005}. 
For $U < U_{c}$, we obtain $C = \pm 1$, indicating a topologically nontrivial phase stabilized by spontaneous time-reversal-symmetry breaking. 
This quantized value persists over a finite interaction window, reflecting the robustness of the QAH state. 
Beyond the critical strength $U_{c}$, the Chern number abruptly drops to $C = 0$, signaling a transition into a topologically trivial SNI.  

The corresponding QAH order parameter, displayed in Fig.~\ref{Chern_no}(b), remains finite in the QAH region ($U < U_{c}$) and vanishes in the SNI phase. 
The simultaneous disappearance of both the Chern number and the QAH order parameter at $U_{c}$ highlights the direct competition between interaction-driven topological order and symmetry-breaking nematic order. 
These results clearly demonstrate a correlation-induced topological phase transition stabilized in an intermediate interaction regime.

\bibliographystyle{apsrev4-2}
\bibliography{references}

\end{document}